\RequirePackage{fix-cm}
\documentclass{svjour3}                     
\smartqed  
\usepackage{graphicx}
\usepackage{amsmath}
\usepackage{amssymb}
\usepackage{booktabs}
\usepackage{epstopdf}
\usepackage{dcolumn}
\usepackage{float}
\usepackage[normalem]{ulem}
\usepackage{color}
\usepackage{dcolumn}
 \usepackage{multirow}
\usepackage{subfigure}
\usepackage{cancel}

\newcommand{\be}{\begin{equation}}

\newcommand{\ee}{\end{equation}}
\newcommand{\bea}{\begin{eqnarray*}}

\newcommand{\eea}{\end{eqnarray*}}

\newcommand{\pup}{p^\uparrow}

\newcommand{\bfp}{\mbox{\boldmath $p$}}

\def\lsim{\mathrel{\rlap{\lower4pt\hbox{\hskip1pt$\sim$}}\raise1pt\hbox{$<$}}}
\def\gsim{\mathrel{\rlap{\lower4pt\hbox{\hskip1pt$\sim$}}\raise1pt\hbox{$>$}}}
\begin{document}
\title{Low-virtuality leptoproduction of open-charm as a probe of the gluon Sivers function}
\author{Rohini M. Godbole         \and
Abhiram Kaushik \and Anuradha Misra
}


\institute{Rohini M. Godbole\at
              Indian Institute of Science \\
              \email{rohini@iisc.ernet.in}           
           \and
           Abhiram Kaushik\at
              Indian Institute of Science \\
              \email{abhiramk@iisc.ernet.in}           
   \and
           Anuradha Misra\at
              University of Mumbai\\
              \email{misra@physics.mu.ac.in}   
}

\date{Received: date / Accepted: date}

\maketitle

\begin{abstract}
We propose low-virtuality leptoproduction of open-charm, $p^\uparrow l\to D^0+X$, as a probe of the gluon Sivers function (GSF). At leading-order, this process directly probes the gluon content of the proton, making detection of a trasverse single-spin asymmetry in the process a clear indication of a non-zero GSF. Considering the kinematics of the proposed future Electron-Ion Collider (EIC), we present predictions for asymmetry using fits of the GSF available in literature. We also study the asymmetry at the level of muons produced in $D$-meson decays and find that the asymmetry is preserved therein as well.

\end{abstract}

\section{Introduction}
\label{intro}
Sivers function is a transverse momentum dependent parton distribution function (TMD) which encodes the correlation between the azimuthal anisotropy in the transverse momentum distribution of an unpolarised parton and the spin of its parent hadron~\cite{Sivers:1989cc,Sivers:1990fh}, $\Delta^N f_{a/p^\uparrow}(x,\mathbf{k}_\perp)\equiv\hat f_{a/p^\uparrow}(x,\mathbf{k}_\perp)-\hat f_{a/p^\downarrow}(x,\mathbf{k}_\perp)$.
In collisions of transversely polarised nucleons off unpolarised nucleons (or leptons), this anisotropy can lead to an azimuthal anisotropy in the distribution of the inclusive final state, i.e a single-spin asymmety (SSA). The SSA for an inclusive process $A^\uparrow B\to C+X$ is defined as
\be
A_N=\frac{d\sigma^\uparrow-d\sigma^\downarrow}{d\sigma^\uparrow+d\sigma^\downarrow}
\ee
where $d\sigma^\uparrow$ and $d\sigma^\downarrow$ represent the cross-section for scattering of a transversely polarized hadron A off an unpolarized hadron (or lepton) B with A being polarised upwards and downwards respectively, with respect to the production plane. 

One of the two main theoretical approaches to discuss these asymmetries is based on factorisation in terms of a hard-part and transverse momentum dependent parton distribution functions and fragmentation functions. While TMD factorisation has only been formally established for two-scale processes, a lot of work has been done on a TMD description of single hard-scale processes under the assumption of factorisation, in what is generally referred to as the generalised parton model (GPM) approach~\cite{DAlesio:2004eso,DAlesio:2007bjf}. In this work, we study the low-virtuality leptoproduction ($Q^2\approx0$) of open-charm as a possible probe of the poorly understood gluon Sivers function (GSF), adopting the GPM framework. At the leading-order (LO) of this process, the production of open-charm happens only via the photon-gluon fusion (PGF) process, making detection of a SSA in this process a direct indication of a non-zero GSF.

In Section 2, we present the parametrisation of the TMDs that we have used. In Section 3, we present the expressions for the SSA in $p^\uparrow l \to D+X$ as well as the results.


\section{Formalism and parametrisation of the TMDs}
\label{sec:1}
The denominator and numerator of the asymmetry (Eq. 2) are given by,
\bea
 d\sigma ^\uparrow + d\sigma ^\downarrow &=& 
\frac{E_D \, d\sigma^{p^\uparrow l \to DX}} {d^{3} \bfp_D} +
\frac{E_D \, d\sigma^{p^\downarrow l \to DX}} {d^{3} \bfp_D}
 = \>
2\int dx_g \, dx_\gamma  \, dz \, d^2 \mathbf{k}_{\perp g} \, d^2 \mathbf{k}_{\perp \gamma} \, 
d^3 \mathbf{k}_{D} \, 
\delta (\mathbf{k}_{D} \cdot \hat{\bfp}_c) \, 
 \> \\
&& \hspace*{-2.0cm} \times ~
{\mathcal C}(x_g,x_\gamma,z,\mathbf{k}_D)~f_{g/p}(x_g,\mathbf{k}_{\perp g}) \>  f_{\gamma/l}(x_\gamma, 
\mathbf{k}_{\perp \gamma}) ~
 \frac{d \hat{\sigma}^{g\gamma \to c \bar c}}
{d\hat t}D_{D/c}(z,\mathbf{k}_D) ~\delta (\hat s +\hat t +\hat u - 2m_c^2)
\label{denominator}
\eea
and
\bea
d\sigma ^\uparrow - d\sigma ^\downarrow &=& 
\frac{E_D \, d\sigma^{p^\uparrow l  \to DX}} {d^{3} \bfp_D} -
\frac{E_D \, d\sigma^{p^\downarrow l\to DX}} {d^{3} \bfp_D}  
 = \>
\int dx_g \, dx_\gamma  \, dz \, d^2 \mathbf{k}_{\perp g} \, d^2 \mathbf{k}_{\perp \gamma} \, 
d^3 \mathbf{k}_{D} \, 
\delta (\mathbf{k}_{D} \cdot \hat{\bfp}_c) \, 
 \>  \\
&& \hspace*{-2.0cm} \times~ {\mathcal C}(x_g,x_\gamma,z,\mathbf{k}_D)~ \Delta ^N f_{g/\pup}(x_g,\mathbf{k}_{\perp g}) \>  f_{\gamma/l}(x_\gamma, 
\mathbf{k}_{\perp \gamma}) ~
\frac{d \hat{\sigma}^{g\gamma \to c \bar c}}
{d\hat t} \>
 D_{D/c}(z,\mathbf{k}_D)~\delta (\hat s +\hat t +\hat u - 2m_c^2).
\label{numerator}
\eea
For the unpolarised TMD PDF we use the standard Gaussian form:\,
\be
f_{g/p}(x,k_\perp;Q)=f_{g/p}(x,Q)\frac{1}{\pi\langle k_\perp^2\rangle}e^{-k_\perp^2/\langle k_\perp^2\rangle},
\ee
with $\langle k_\perp^2\rangle=0.25$ GeV$^2$. For the density of quasi-real photons in a lepton, we use a similar Gaussian form as well, with the Weiszacker-Williams distribution for the collinear part and a Gaussian $k_\perp$-spread of width $\langle k_{\perp \gamma}^2\rangle=0.1$ GeV$^2$.
We also take the transverse-momentum-dependence of the FF to be Gaussian with a width $\langle k_{\perp D}^2\rangle=0.25$ GeV$^2$.
For the Sivers function, 
we use the parametrization ~\cite{DAlesio:2015fwo}
\bea
\Delta^N f_{g/p^\uparrow}(x,k_\perp;Q)=2\mathcal{N}_{g}(x)f_{g/p}(x,Q)~ \frac{\sqrt{2e}}{\pi} \sqrt\frac{1 - \rho}{\rho} {k_\perp} 
\frac{e^{- k_{\perp}^2 / \rho \langle  k_{\perp}^2 \rangle}}{{\langle  k_{\perp}^2 \rangle}^{3/2}},
\eea
where $0<\rho<1$. $\mathcal{N}_g(x)$ here parametrises the $x$-dependence of the GSF and is generally written as
\begin{equation}
\mathcal{N}_g(x)=N_g x^{\alpha_g}(1-x)^{\beta_g}\frac{(\alpha_g+\beta_g)^{\alpha_g+\beta_g}}{\alpha_g^{\alpha_g} \beta_g^{\beta_g}}.
\end{equation}
The requirement that the Sivers function satisfy the positivity bound $|\Delta^Nf_{g/p^\uparrow}(x,\mathbf{k}_\perp)|/2f_{g/p}(x,\mathbf{k}_\perp)\leq1$ $\>\forall \>x, \mathbf{k}_\perp$, implies $|\mathcal{N}_g(x)|<1$.

In this work, in order to demonstrate the efficacy of the suggested probe, we explore two choices for the gluon Sivers function: 
\begin{enumerate}
\item the Sivers function with the positivity bound saturated, viz., $\mathcal{N}_g(x)=1$ and $\rho=2/3$.
\item the SIDIS1 and SIDIS2 extractions of the gluon Sivers function from Ref.~\cite{DAlesio:2015fwo}, which have been obtained using data on mid-rapidity pion production measured by the PHENIX experiment at RHIC~\cite{Adare:2013ekj}.
\end{enumerate}
The first choice, which we call the `saturated' Sivers function, would give an upper bound on the magnitude of the asymmetry for a fixed width $\langle k^2_\perp\rangle$ and $\rho$, and for a given choice of unpolarised gluon density. The parameter $\rho$ is set to $2/3$ in order to maximize the first $k_\perp$-moment of the Sivers function, following Ref.~\cite{DAlesio:2010sag}.

The SIDIS1 and SIDIS2 GSFs from Ref.~\cite{DAlesio:2015fwo} are the first (and so far, only) available extractions of the GSF in a GPM framework. They were obtained by fitting to the PHENIX data on $A_N$ for inclusive pion production in the midrapidity region at RHIC. In their analysis, they used quark Sivers functions extracted from semi-inclusive deep inelastic scattering data to account for the quark contribution to the asymmetry, $A_N$. The two GSF extractions differ in the choice of QSFs, as well as that of the fragmentation functions adopted in the fitting process. As a result, they show very different $x$-dependencies, with SIDIS1 being larger in the moderate-$x$ region and SIDIS2 being larger in the low-$x$ region. The fact that these widely different choices for the GSF are consisten with the same data on $A_N$ underscores the utility of the process proposed by us for determination of the GSF. The values of the parameters of the two GSF fits are given in Table I.

\begin{table*}[t]
\centering
\begin{tabular}{|l|l|l|l|l|l|l|}
\hline
SIDIS1 & \multicolumn{2}{l|}{$N_g=0.65$} & $\alpha_g=2.8$ & $\beta_g=2.8$ & $\rho=0.687$ & \multirow{2}{*}{$\langle k^2_\perp\rangle=0.25$ GeV$^2$} \\ \cline{1-6}
SIDIS2 & \multicolumn{2}{l|}{$N_g=0.05$} & $\alpha_g=0.8$ & $\beta_g=1.4$ & $\rho=0.576$ &                                                        \\ \cline{1-7}
\end{tabular}
\caption{Parameters of the GSF fits from Ref.~\cite{DAlesio:2015fwo}.}
\label{SIDIS-gluon-fits}
\end{table*}

\section{Results}

\begin{figure*}[t]
\begin{center}
\vspace*{-1cm}
\includegraphics[width=0.8\linewidth]{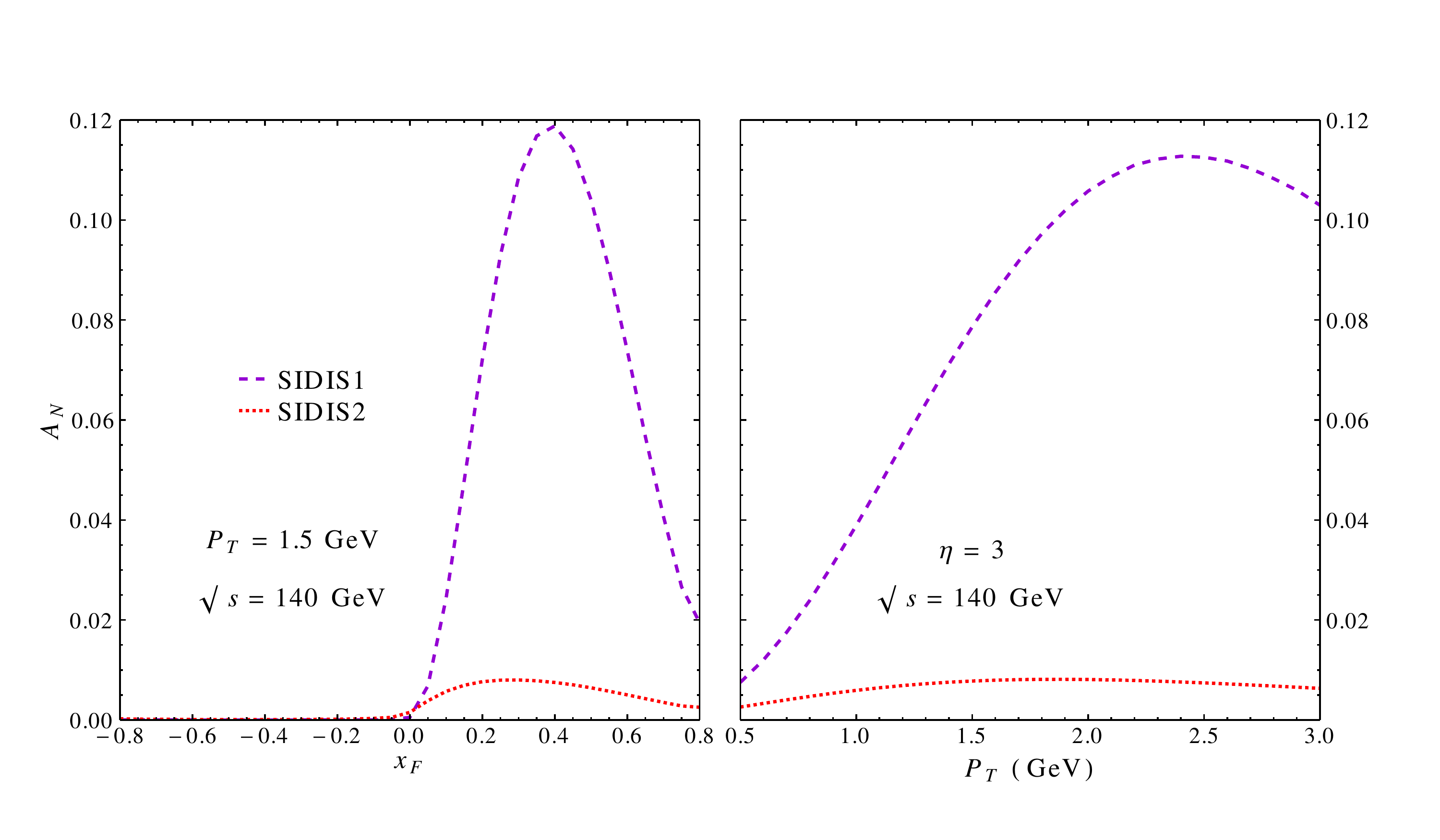}
\vspace*{-0.5cm}
\caption{SSA from GSF fits of Ref.~\cite{DAlesio:2015fwo} at EIC as a function of $x_F$ (at fixed $P_T$, left panel) and $P_T$ (at fixed $\eta$, right panel). Using MRST2001LO PDF for collinear gluon density. Figure from Ref.~\cite{Godbole:2017fab}.}
\label{EICsidisAN}
\end{center}
\vspace*{-0.5cm}
\end{figure*}

Fig.~\ref{EICsidisAN} we show the asymmetries obtained using the SIDIS1 and SIDIS2 fits~\cite{DAlesio:2015fwo}. Since the fits were obtained using MRST2001LO  PDFs~\cite{Martin:2001es} for the collinear densities, to be consistent, we use the same. Both fits give asymmetries much smaller than allowed by the positivity bound. Further the SSAs for SIDIS1 and SIDIS2 differ from each other substantially and thus offer discrimination between the two GSF extractions. While we have not shown the plots here, we find that the probe is able to discriminate between the two fits at COMPASS kinematics as well~\cite{Godbole:2017fab}.

\begin{figure*}[t]
\begin{center}
\vspace*{-1cm}
\includegraphics[width=0.8\linewidth]{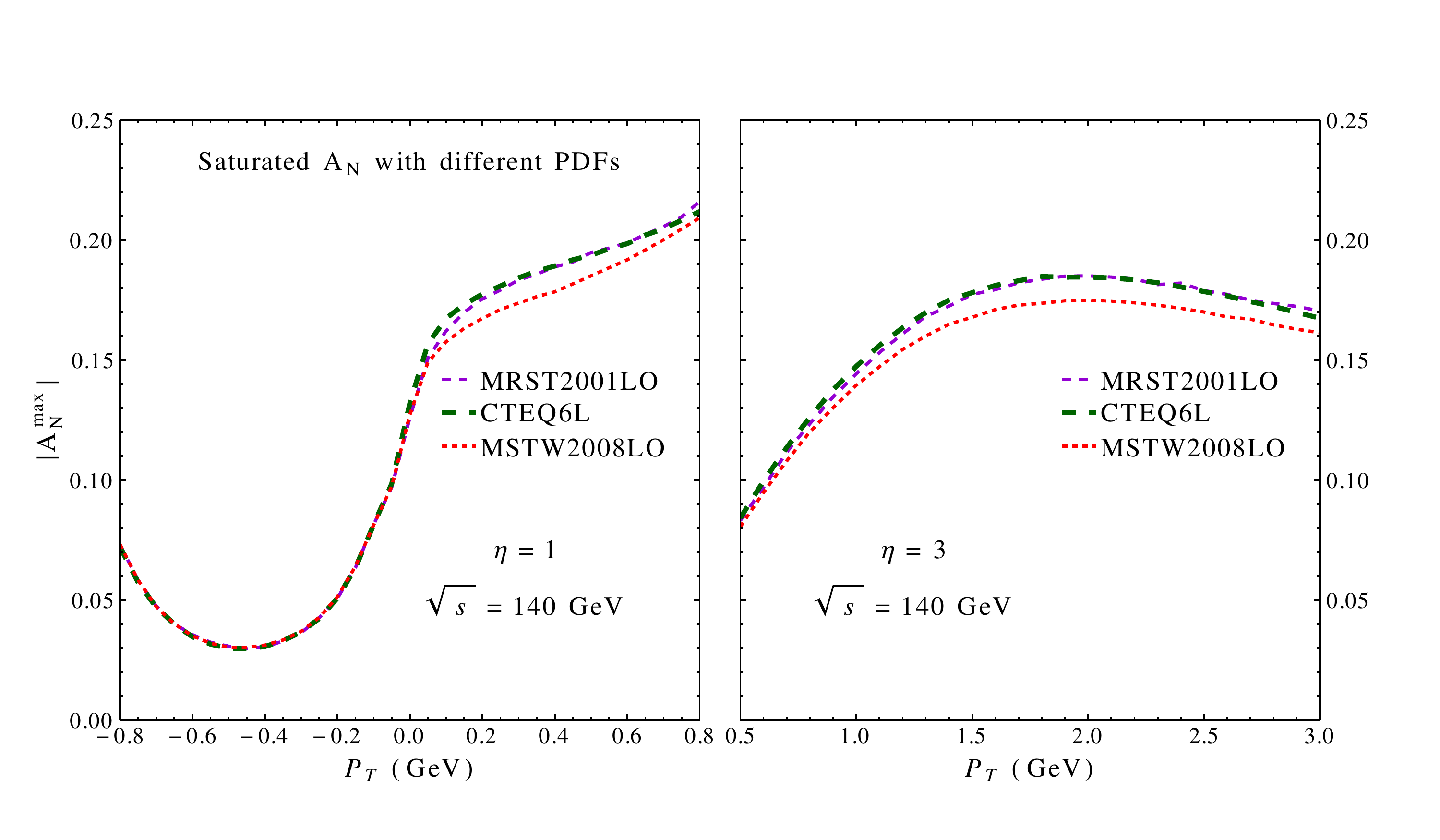}
\vspace*{-0.5cm}
\caption{Variation of results for saturated GSF for different choices unpolarised gluon densities. We consider the MRST2001LO (green, long-dashed), CTEQ6L (purple, short-dashed) and MSTW2008LO (red, dotted) gluon distributions. }
\label{EICsatAN}
\end{center}
\vspace*{-0.7cm}
\end{figure*}

In Fig.~\ref{EICsatAN}, we show estimates for the maximum value of the magnitude of the asymmetry $|A_N|$ at the Electron-Ion Collider (EIC), calculated by using the saturated gluon Sivers function. In case of the saturated GSF, the $x$-dependence is determined only by the choice of unpolarised gluon densities. Therefore, in  order to demonstrate effects that uncertainties in the gluon densities might have on the probe, we have presented results for three different choices of leading-order (LO) unpolarised PDFs, MRST2001LO, CTEQ6L~\cite{Pumplin:2002vw} and MSTW2008LO~\cite{Martin:2009iq}. We find that the results are somewhat affected by the choice of PDF set, with the estimate for the saturated asymmetry varying by up to 6\% between CTEQ6L and MSTW2008. In general, with the large centre of mass energy of the EIC, the general features of $|A^\text{max}_N|$ are similar to those that had been observed in calculations for $pp$ collisions at RHIC~\cite{Anselmino:2004nk,Godbole:2016tvq}, especially the azimuthal suppression of the asymmetry in the backward hemisphere ($x_F<0$).

\begin{figure}[h]
\begin{center}
\vspace*{-1cm}
\includegraphics[width=0.5\linewidth]{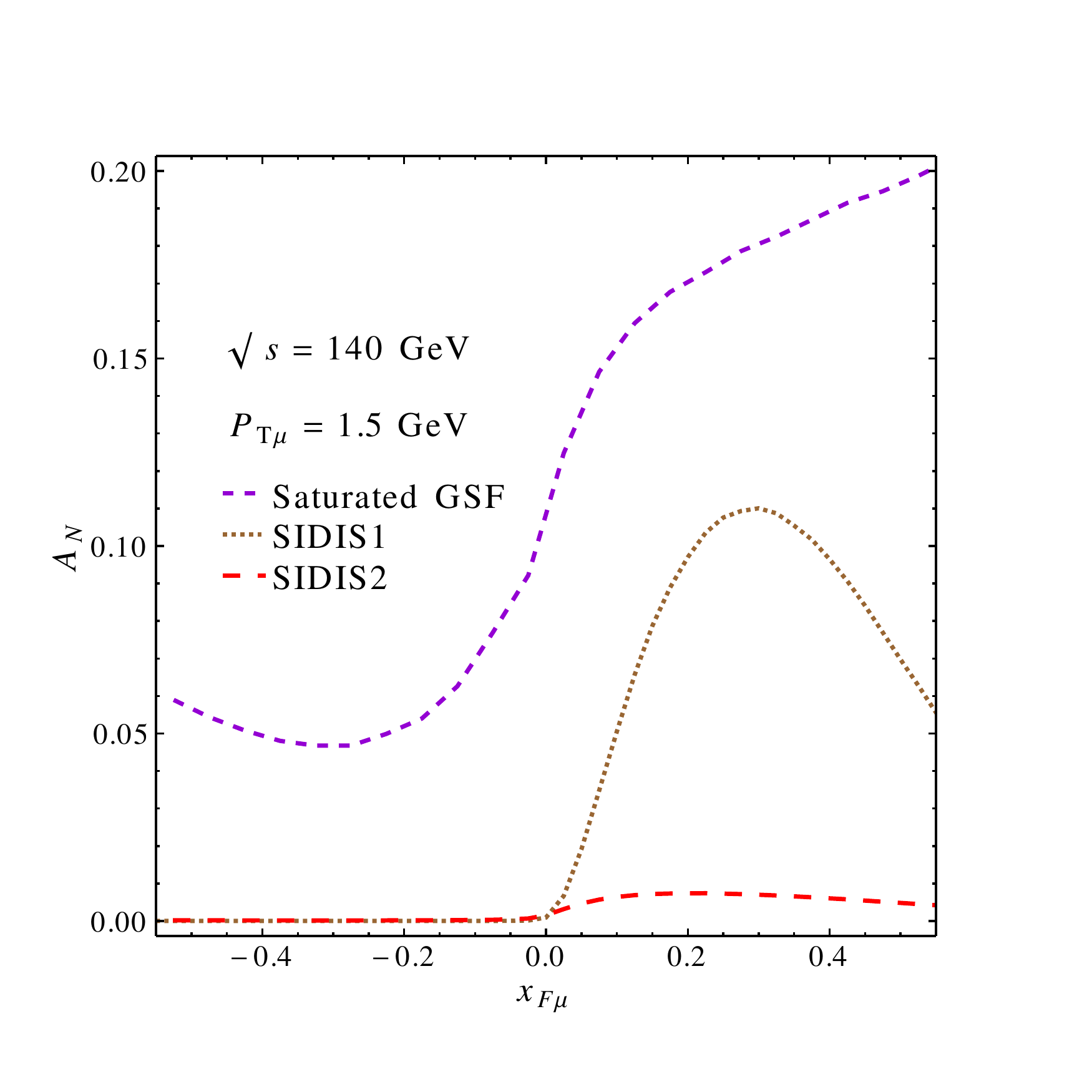}
\vspace*{-0.5cm}
\caption{SSA for decay-muons. Using MRST2001LO PDF for collinear gluon density. Figure from Ref.~\cite{Godbole:2017fab}.}
\label{EICsatANdecay}
\end{center}
\vspace*{-0.5cm}
\end{figure}

Since the experiments detect $D$-mesons through the muons produced in the decay, it is an interesting question to ask, how much -- if any -- of the SSA present at the level of the $D$-mesons is transmitted at the level of the detected muons? This has the advantage that the asymmetry measurement will not have the additional errors due to $D$-meson reconstruction. The results for the SSA in the kinematics of the decay muons are presented in Fig.~\ref{EICsatANdecay} as a function of ${x_F}_\mu=2{P_L}_\mu/\sqrt{s}$, with the muon transverse momentum ${P_T}_\mu=1.5$ GeV. It appears that an azimuthal anisotropy in $D$ production would be retained significantly in the decay-muons. Peak values of the muon $A^\mu_N$ for all three choices of the GSF are close to those obtained for the meson.

\vspace*{-1.5cm}

\section{Conclusions}
We find that an asymmetry of upto around 22\% is allowed by the saturated gluon Sivers function at the EIC. Further,  asymmetry is significantly retained in the distribution of decay muons. We also find that the probe is able to discriminate well between the two available phenomenological  fits of the gluon Sivers function, both of which were obtained using the same data on $A_N$ measured at PHENIX. Thus we see that this process offers a good probe of the gluon Sivers function and can be of help in a global extraction of it.

\begin{acknowledgements}
R.M.G. wishes to acknowledge support from the Department of Science and
Technology, India under Grant No. SR/S2/JCB-64/2007 under the J.C. Bose Fellowship scheme.  A.M would like to thank the Department of Science and Technology, India for financial support under Grant No.EMR/2014/0000486. A.M would also like to thank the Theory Division, CERN, Switzerland for their kind hospitality.
\end{acknowledgements}

\vspace*{-0.5cm}

\bibliographystyle{spphys}       

\end{document}